\documentclass[sigconf]{acmart}

\usepackage{todonotes}

\newenvironment{packed_itemize}{
\begin{list}{\labelitemi}{\leftmargin=2em}
\vspace{-6pt}
 \setlength{\itemsep}{0pt}
 \setlength{\parskip}{0pt}
 \setlength{\parsep}{0pt}
}{\end{list}}

\AtBeginDocument{%
  }

\setcopyright{cc}
\copyrightyear{2023}
\acmYear{2023}

\acmConference[ACSACARTCOMP '23]{
	Cybersecurity Artifacts Competition and Impact Award
  }{December 04--08,
  2023}{Austin, TX}

\begin{document}

\title{Zipr:  A High-Impact, Robust, Open-source, Multi-platform, \\ Static Binary Rewriter}

\author{Jason D. Hiser, Anh Nguyen-Tuong, Jack W. Davidson}
\email{{hiser,jwd,an7s}@virginia.edu}
\affiliation{%
  \institution{University of Virginia}
  \streetaddress{85 Engineer's Way}
  \city{Charlottesvile}
  \state{Virginia}
  \country{USA}
  \postcode{22904}
}

\renewcommand{\shortauthors}{Hiser et al.}

\begin{abstract}

Zipr is a tool for static binary rewriting, first published in 2016.  
Zipr was engineered to support arbitrary program modification
with an emphasis on low overhead, robustness, and flexibility to perform
security enhancements and instrumentation.  Originally targeted to Linux x86-32
binaries, Zipr now supports 32- and 64-bit binaries for X86, ARM, and MIPS architectures,
as well as preliminary support for Windows programs.

These features have helped Zipr make a dramatic impact on research.
It was first used in the DARPA Cyber Grand Challenge to take second place overall,
with the best security score of any participant, Zipr has now been used
in a variety of research areas by both the original authors as well as third parties.
Zipr has also led to publications in artificial diversity, program instrumentation, program repair, 
fuzzing, autonomous vehicle security, research computing security, as well
as directly contributing to two student dissertations.  
The open-source repository has accepted accepted patches from several external authors,
demonstrating the impact of Zipr beyond the original authors.

\end{abstract}



\received{2 October 2023}

\maketitle

\section{Introduction}


The impetus for creating Zipr was the need to apply many patches to arbitrary binaries efficiently.  
We had a static analysis system for binaries that could identify potential buffer overflows and recommend
patches, and we needed to determine the runtime costs of applying the patches.  
While we had access to dynamic binary translators, runtime and memory overheads were not appropriate for 
the embedded systems we were targeting.~\cite{scott2001strata, scott2003retargetable, hiser2007evaluating}
At the time, there were two flavors of static binary rewriters: Ones that kept a copy of the original program text to 
deal with disassembly errors and ones that overwrote the code to be instrumented with a trampoline
to an unused address to execute the instrumented code snippet, then return to the original code 
to continue execution.

Keeping a full second copy of the program was untenable for embedded systems.  Further,
we knew that our static analyzer likely had many false positives (as static analyzers tend to do), 
so we did not want our instrumentation to suffer from the cache, branch predictor, and memory penalty
overheads of frequent trampolining inside time-critical kernel loops.  
We needed a rewriting system suitable for embedded systems, and nothing available
met the requirements:  1) low overhead (in terms of memory and performance), 2) robust for a large range of common programs,
and 3) the ability to cheaply invoke arbitrary instrumentation for any instruction in the program.
Of course, these features are desirable in most environments but absolutely necessary for embedded systems. 

We realized that instead of trampolining for patched instructions, we could trampoline only the (comparatively infrequent)
indirect branch instruction targets.  This approach allows instrumentation of any instruction.  Zipr works by putting a trampoline 
at each indirect branch target, and then places the remaining code blocks around these trampolines.  
To save space, Zipr lays out blocks between trampolines with a
best-fit algorithm, and often simply replaces the trampoline with the correct code, avoiding any overhead at all.

The primary publications describing Zipr were published in 2016 and 2017~\cite{davidson2016system,hawkins2017zipr,hiser2017zipr++,hawkins2017securing}.
The open source release~\cite{ziprsrc} occurred in 2019 after a dependency on a commercial software package, IDA Pro~\cite{eagle2011ida}, was eliminated.   
Zipr has since been extended for multiple platforms and demonstrated effective, robust binary rewriting.  
Section~\ref{sec:current} discusses the Zipr artifact's current release, while Section~\ref{sec:contributions}
describes the impacts that Zipr has made in artificial diversity, program instrumentation, program repair,
fuzzing, autonomous vehicle security, research computing security, as well
as directly contributing to two student dissertations and numerous papers in key areas of security and privacy.



\vspace{-.1in}
\section{The Zipr Artifact}
\label{sec:current}

\paragraph{Zipr Architecture}
The Zipr architecture includes a front end that parses binary programs and lifts it into a low-level Zipr intermediate representation (IR).
This IR is stored in a database we call the IRDB.  
The front end detects instructions, functions, data objects, indirect branch targets, and stack unwinding information (also called exception handling (EH) tables)
and deposits this information in the IRDB.
The IRDB contains information about control transfers, data object locations, and whether the IR is complete enough to \emph{unpin}
a data object or instruction, etc.  One key feature of the IRDB storage mechanism is that it allows for uncertain and conflicting information.
For example, our disassembly technique allows multiple disassemblers to union their results.  Infrequently, disassemblers disagree on the correct interpretation
of a sequence of bytes.  The IRDB stores \emph{both} representations in the IRDB, and ultimately, both interpretations will end up in the final program (assuming
no user-specified transformations remove them).  While only one interpretation is correct, the program semantics will be preserved because
only one of them will ever execute dynamically.
The incorrect sequence will result in extra memory and disk space usage, but will not affect correctness.

After building the IR, the user can select one or more composable transformations to the IR.  
To ease the task of writing transformations, Zipr provides an SDK for reading, writing, and modifying the IRDB.~\cite{irdbsdk}
The current SDK is written for C++ programming, but it has always been envisioned that one could directly modify the IRDB 
using SQL or write interfaces for other languages.  

Besides direct access to any program item, we recognized that many transformations would want to analyze the program in some way before
making a transformation.  To facilitate this approach, the SDK provides common analysis techniques, such as the ability to create a control flow 
graph (CFG) for a function; the dominator graph for the CFG, a call graph of functions in the IR; the dead registers at each program point;
and the disassembly and set/used registers for each instruction.  While the set of analyses is by no means complete, the SDK is 
easily extendable by writing a new transformation and compiling it into a dynamically-linked library that others can reuse.

To help potential users write Zipr transformations, we have provided a variety of example transformations:
\begin{packed_itemize}
\item An IRDB Cookbook~\cite{irdbcookbook} with well-documented transformations provides documentation on writing a plugin and includes three example plugins.
\texttt{initialize\_stack} inserts code at the beginning of every function to zero-initialize the stack, preventing exploits relating to uninitialized stack variables.  
This example shows how to iterate over functions, inspect a function's stack frame, and insert instrumentation. 
\texttt{kill\_deads} runs the IRDB SDK's dead register analysis and inserts a set for every dead register detected at every instruction in the program.  
This example shows how to invoke IRDB SDK analyses and act on the results, while simultaneously being a surprisingly effective test of the analysis results.
\texttt{stack\_stamp} modifies function entry and exit code to XOR the program's return address with a randomly selected value.  Besides being a potential security enhancement, this example shows how to iterate over functions and modify each function's entry and exit points.
\item Zafl is an instrumentation pass built on Zipr and the IRDB SDK to instrument a program with AFL-style instrumentation.~\cite{zafl,nagy2021breaking,wong2022american,fioraldi2020aflpp}.  
Zafl yields an instrumented program that is statistically similar to a program with compiler-inserted instrumentation in terms of bug-finding abilities.
\item The \texttt{p1transform} transformation randomizes the amount of stack space each function uses, providing an example of how to modify stack frames for a given function and providing artificial diversity for the transformed program.~\cite{p1transform}
\end{packed_itemize}

After completing the (possibly empty) list of user-specified transformations, Zipr invokes the back end to reconstitute the IR directly into
an executable program.  As we recognized that users may want to control the final layout of the program, a second SDK is provided, the Zipr Backend SDK.~\cite{ziprsdk}
The primary purpose of this SDK is to allow plugins to the reconstitution engine to allow the user to control basic block placement, and apply
user-specified relocations.  A Zipr relocation is analogous to a linker relocation in the compiler world.  Built-in relocations for changing, for example,
a data item to point at an instruction, are included.
The Selective CFI transformation ~\cite{scfi} demonstrates how to use the Zipr Backend SDK (described below) to place executable or not-executable nonces in the code
to implement a highly effective version of CFI directly on binary programs.

To facilitate quick adoption by new users, the Zephyr Gitlab repository includes docker images with pre-built Zipr installations, ready to run with a single command.

\paragraph{Zipr Platforms}


Zipr's most robust architecture is the x86/64 Linux platform.   
This platform supports all common compilers (gcc/g++, icx, clang/clang++ and the obfuscating LLVM 
compiler, OLLVM).~\cite{gough2004introduction,icx,lattner2008llvm,junod2015obfuscator},
all common compiler optimization levels and flags ( \texttt{-O0},
\texttt{-O1}, \texttt{-O2}, \texttt{-O3}, \texttt{-Os}, \texttt{-Ofast},
\texttt{-fomit-frame-pointer}, \texttt{-static}, \texttt{-fPIC} \texttt{-fPID} \texttt{-pie}, etc.),
and most languages that are commonly compiled (C, C++, Fortran, Ada, Rust, including variants of said languages 
such as C++11, C++17, Fortran99, etc.)  Notably missing is GoLang support due to a currently undiagnosed race condition with the GoLang custom exception handling format.  
Both static and dynamically linked programs and shared libraries are supported.
The platform's robustness is demonstrated in our test suite
for Zipr testing, which contains thousands of binaries generated from programs used in industry.~\cite{zipr_x86_testing}  
X86/32 code, though the first target of Zipr, has yet to achieve as much attention as we have had fewer partners working
with that code base.

The next most supported platform is 32-bit ARM on Linux.  We have automated regular regressions for coreutils
and several other programs used in common software deployments.~\cite{zipr_arm_testing}  While there are known bugs, many large programs
work and demonstrate the approach's feasibility for this platform.  

We also have preliminary support for 64-bit ARM code and 32- and 64-bit MIPS code on the Linux platform.  While we do 
not yet have a robust test suite with widely-used programs, a variety of coreutils programs have demonstrated successful transformations.
Preliminary support is also available for Windows 32- and 64-bit executables.

%
%
%
%
%

\vspace{-.1in}
\section{Zipr Impacts}
\label{sec:contributions}

\vspace{-.05in}
\subsection{Commercial Adoption}

\paragraph{GrammaTech}

GrammaTech was one of the first commercial adopters of the Zipr technology with the authors' collaboration.
GrammaTech concluded that Zipr was a good fit for their entry into the
DARPA Cyber Grand Challenge (CGC).  CGC was a fully-autonomous capture the flag (CTF)
competition.   Each team built a Cyber Reasoning System (CRS), and the CRSes competed
to autonomously protect their (vulnerable) services and exploit
vulnerabilities in competitors' services.  A CRS scored points for successfully
defending against an exploit, but was harshly penalized for increases in
runtime, on-disk space, or memory usage. 
Thus, Zipr was a natural fit due to its low runtime overhead, minimal on-disk expansion,
and modest memory footprint.~\cite{nguyen2018xandra}  

The primary defensive mechanism was a control flow integrity (CFI) implementation~\cite{burow2017control,abadi2009control}, which is now open source.~\cite{scfi}
The CFI implementation, dubbed Selective CFI, leveraged the IRDB analyses to detect indirect branches that were safe from exploitation,
and elide expensive instrumentation in those cases to help minimize scoring penalties for using excess resources.
Zipr and the Selective CFI implementation achieved the \#1 score for defeating exploits created by the other competitors.

Later, Zipr's exception handling parser~\cite{hiser2017zipr++} was adopted by GrammaTech.  
They were able to re-use Zipr's \texttt{libEHP}~\cite{libehpsrc} component in their Datalog Disassembler (\texttt{ddisasm})\cite{flores2020datalog}.
GrammaTech's work also helped improve \texttt{libEHP} by submitting source code updates with bug fixes, build system improvements, and
porting it to operate on MS Windows platforms.

Finally, ONR sponsored GrammaTech to perform a broad comparison of binary rewriters, which includes both Zipr and \texttt{ddisasm}.~\cite{schulte2022broad}
Zipr performed very well.
It was the only binary rewriter to transform every program in the extensive test suite successfully.
Zipr and \texttt{ddisasm} were the only binary rewriters that were able to transform a majority of programs, while many other
rewriters were deemed ``too academic'' for wide adoption.

\paragraph{Dependable Computing}

A joint project between Dependable Computing and the authors leveraged formal verification techniques
to meet security and privacy needs for embedded systems where development artifacts such as source code were unavailable.~\cite{davidson2016system}
Zipr was critical for meeting embedded systems' strict overhead limits, and was able to apply patches to meet
security requirements.  Overall, Zipr demonstrated that it could rewrite a binary with patches and yield less than 5\%
performance penalty.

\paragraph{Apogee Software}

Apogee software, as part of the DARPA Cyber Fault-tolerance Attack Recovery (CFAR) project, worked to realize DARPA's vision 
of executing multiple diverse copies of a program in parallel to detect security violations.~\cite{co2016double} 
The key insight is that many attacks must be customized to the low-level details of a program under attack.  
For example, an attack might include an absolute address or relative
offset to exploit the program.  If the same input is sent to multiple diverse copies of the program,
the programs will behave differently, and that difference can be detected and used to invoke recovery mechanisms.

In collaboration with the UVA team, Zipr was used to generate artificial diversity for key web server programs such as
Apache and NginX.~\cite{laurie2003apache,soni2016nginx}
Zipr was used to randomize code, data layouts, and locations for the stack, heap, and global address spaces.  
The Zipr backend SDK was used to control
code and data layout locations so that detection of certain classes of attacks was provably certain.
For example, Zipr could create two variants of a program such that their address spaces were disjoint.
Thus, any attack that injected an absolute address would crash at least one variant.

Other teams in the program did similar transformations using compiler technology instead of operating directly on the binary.
As such, when an Ada program was introduced, the Zipr transformer was the only tool that successfully
produced diverse variants for parallel execution.  In fact, thanks to the robustness of Zipr binary analysis, little additional effort
was required when the source language changed.

Zipr was also leveraged to enact automatic program repair.~\cite{jones2019defeating}  The program repair worked by instrumenting  a program's
input points to memorize recent program inputs.  If the input subsequently caused a security violation, the input could be marked
as malicious.  Future attempts to feed the input to the program caused the input to be ignored and not processed by the program.
This technique helps defeat denial-of-service (DOS) attacks where the same input causes the program to repeatedly crash and restart, preventing
legitimate requests from being serviced while the program is busy restarting.


\paragraph{Red Hat}

Red Hat and the authors collaborated to deploy several Zipr-based tools for ensemble fuzzing with an unnamed collaborator.~\cite{zafl,turbo}
Unfortunately, as is often the case with commercial partners, we can provide no additional detail due to confidentiality restrictions.


\vspace{-.1in}
\subsection{Student Research}

Zipr has provided direct, major contributions to two Ph.D. student dissertations.

\paragraph{Dr. William Hawkins}

The first student dissertation we will discuss is William Hawkins's dissertation who was advised by Dr. Jack W. Davidson.~\cite{hawkins_dissertation}
Dr. Hawkins's dissertation, granted by the University of Virginia, included content from several of the original Zipr publications~\cite{hawkins2017zipr,hiser2017zipr++}
but also included several security-enhancing transformations.  

One of Dr. Hawkins's dissertation's contributions was a technique called dynamic canary randomization.~\cite{hawkins2016dynamic}
The technique was designed to walk the stack periodically, find the stack canaries used, and update the canaries to a new value.
The primary goal of the technique was to thwart attacks where an attacker can leak stack canaries and then later leverage the
leaked value to attack critical control flow data on the stack.

The second of Dr. Hawkins's contributions was a system called Mixr.\cite{hawkins2017mixr}
Mixr was a tool that leveraged a Zipr plugin to lay out the program into fixed-size code blocks.
Each block had metadata about locations within the block that would need to be patched if the code were to move.
When the transformed program was executed, user-specified mechanisms triggered randomization of the code locations.
Mixr would randomly swap a predetermined number of code blocks, leveraging the metadata to update code offsets,
and walk the execution stack to update return addresses as needed.  Mixr provided a \emph{moving target defense},
a mechanism to invalidate information that an attacker may attempt to learn over time, such as code locations.~\cite{lei2018moving}
Dr. Hawkins's work evaluated differences in code block sizes and re-randomization times and policies on the performance
and security of an application.

\paragraph{Stefan Nagy}
Dr. Stefan Nagy and his advisor Dr. Matthew Hicks at Virginia Tech worked on techniques to improve closed-source application fuzzing.~\cite{nagy_dissertation}
Two of Dr. Nagy's major contributions were published in top tier conferences and leveraged the Zipr-based plugin, Zafl.\cite{zafl}

The first paper examines the features needed to achieve compiler-quality fuzzing instrumentation on a binary program.~\cite{nagy2021breaking}
The paper leveraged the IRDB SDK's analyses to add AFL-style instrumentation to a binary program.
The work presents several novel techniques, as well as leveraging IRDB analyses such as dead register analysis to instrument the program
to produce code that has the same performance (in terms of both execution time and bug-finding capabilities) as compiler-generated instrumentation.  
These techniques significantly improved the state of the art in binary-only fuzzing, which previously relied on heavy-weight instrumentation systems such as QEMU.

Dr. Nagy's second paper leverages hardware breakpoints to detect when fuzzing inputs causes new behavior in a program.~\cite{nagy2021same}
When new behaviors are observed, heavier instrumentation determines the new behaviors, and remove the corresponding 
hardware breakpoints before fuzzing continues.
Because most fuzzing inputs do not generate new behaviors, the technique rarely uses the more expensive instrumentation, which amortizes their costs.
This technique ultimately leads to near-zero overhead for tracing fuzzing inputs in a program.
The technique leverages the IRDB SDK's 
loop analysis methods to insert accounting instrumentation code in loop headers to get coverage-preserving, coverage-guided tracing.

\paragraph{Student Information Requests}

We have received numerous requests from students for bug fixes, capability suggestions, etc.  
One example is Franziska Maeckel, a student from the University of Bamberg in Bamberg, Germany.
Mr. Maeckel is working on an open-source version of an address sanitizer for Zafl as part of his dissertation.~\cite{binarymsan,serebryany2012addresssanitizer}
He contacted us regarding a suspected bug in the Zipr layout engine due to the Zipr backend reporting an error in the IR.
We worked with him to track
the issue to his transformation violating one of the IR's invariant properties.

Various other students have contacted us as well, indicating that people are in fact interested in, and using Zipr. 
We have seen social media posts from students elated that Zafl has produced an ``order of magnitude improvement in fuzzing speed'' for their CTF challenge problem.
Unfortunately, as the project is freely available at Zephyr's Gitlab instance, we cannot track the number of anonymous downloads, 
how people may be using the technology, or re-sharing of the project via other means.

\vspace{-.1in}
\subsection{Sponsored Research}


\paragraph{Kevlar}

The Kevlar project was sponsored research by the authors with funding from the Air Force Research Labs (AFRL).
The project's goal was to transition to practice a variety of research tools for randomizing program layouts
and adding hardening transformations to applications.
Many of these transformations were previously
realized using a dynamic rewriter, and the project leveraged the Zipr technology to apply the transformations statically, significantly improving the viability for commercial and government applications
for several reasons:
\vspace{.1in}
\begin{packed_itemize}
\item The Zipr static rewriting technology is more performant and memory frugal, making adoption more likely.
\item The statically rewritten binary can be easily tested in situ, as minimal environmental changes are needed to add security.
Only changing the target binaries is needed, as opposed to additional runtime software, etc.
Such changes are harder to deploy in many settings due to regulatory approval, testing restrictions, or other non-technical issues.
\item Humans need to gain trust in the system, and this trust is much more difficult to gain with dynamic translation systems.
Trust in static translation systems can be audited
for correctness by experts and enhanced with semi-automatic formal verification techniques.
\end{packed_itemize}
\noindent Zipr dramatically improved the practicality of the security techniques.


\paragraph{Trusted and Resilient Computation}

The Trusted and Resilient Mission Operations (TRMO) project, and its follow-on project,
Trusted and Resilient Systems (TRSYS) sought to provide security enhancements for autonomous
vehicles.~\cite{davidson2019trusted,leach2022start}
Myriad Zipr-based plugins provided cyber-attack detection capabilities.  
Program instrumentation relayed detected attacks to a supervisor module, which invoked various forms of program repair.
For example, in an autonomous quad-copter, the attack-response module would put the copter
into ``hover'' mode with a trusted controller while the more capable, 
yet untrusted controller program was being analyzed and repaired.
If the repair is successful, the repaired controller could resume mission operations after a cyber attack.

The Zipr instrumentation system was vital to this work.  It provided the low-overhead
detection capabilities needed to enact program repair, and stop the cyber-attack
from being undetected or immediately fatal.

\paragraph{Helix++}
The Helix++ project aimed at transitioning Zipr-based plugins to practice by leveraging
Docker containers.~\cite{davidson2023helix++, helix++bof}
By protecting applications
in commonly used Docker containers and making these protected containers publicly available, 
we hoped to lower the cost of having developers use Zipr-protected applications. 

We specifically worked with UVA's Research Computing staff to identify key services and build a repository of
hardened Docker containers.
We are in the process of working with them to deploy
their hardened application to end users and hope that this model can be used to lower the cost
of adopting academic security concepts.





\vspace{-.1in}
\section{Conclusions}


This paper presented the Zipr binary rewriter as an artifact, published in 2016 and 2017~\cite{davidson2016system,hawkins2017zipr,hiser2017zipr++,hawkins2017securing}
and open-sourced in 2019~\cite{ziprsrc}.  Zipr's SDKs and plugin architecture target multiple platforms, focusing on efficient binary rewriting to apply diversity and hardening
transformations to improve program security.  Zipr's robust and efficient nature has been leveraged by the authors, 
their collaborators, and third parties to enhance the body of knowledge through publications and dissertations.  Zipr also has a history of
providing open-source tools for diversity, program repair, fuzzing, research computing, and autonomous vehicle security.
We expect that Zipr will continue to be a valuable tool for enhancing program security, and we will continue to make it available and support
user requests for features, bug fixes, and clarifications of included features.

\begin{acks}
This material is based upon work supported by the National Science Foundation under Grant No.  2115130.
\end{acks}

\bibliographystyle{ACM-Reference-Format}
\bibliography{refs}

%

\end{document}